\documentclass[a4paper,fleqn,usenatbib]{mnras}
\usepackage{amssymb,amsmath,graphicx,psfrag,mathtools}
\usepackage[T1]{fontenc} 
\usepackage{ae,aecompl} 
\usepackage{url}
\usepackage{revsymb}
\hypersetup{draft} 
\title[The Environment and Mass of FRB 190520B]{The Environment and Constraints on the Mass of FRB 190520B}
\author[J. I. Katz]{
J. I. Katz,$^{1}$\thanks{E-mail katz@wuphys.wustl.edu} 
\\
$^{1}$Department of Physics and McDonnell Center for the Space Sciences,
Washington University, St. Louis, Mo. 63130 USA 
}
\date{Accepted XXX.  Received YYY; in original form ZZZ} 
\pubyear{2022} 
\date{\today}
\begin{document} 
\label{firstpage} 
\pagerange{\pageref{firstpage}--\pageref{lastpage}} 
\maketitle 
\begin{abstract}
  	Recent observations \citep{A22,D22} of FRB 20190520B have revealed
	rapid fluctuation of its Dispersion Measure within apparently fixed
	bounds, as well as a reversal of its Rotation Measure.  The
	fluctuations of Dispersion Measure are uncorrelated with the
	intervals between bursts, setting upper bounds $\sim 10\,$s on any
	characteristic time scale of the dispersing region; it must be very
	compact.  Measurements of the full dependence of the dispersive time
	delay on frequency may determine the actual electron density and the
	size of this region.  It is possible to set a lower bound on the
	mass of the FRB source from constraints on the size of the
	dispersing region and its time scale of variation.  Comparison of
	the variations of DM and RM leads to an estimate of the magnetic
	field $\sim 500\,\mu$G.
\end{abstract}
\begin{keywords} 
radio continuum, transients: fast radio bursts, accretion, accretion discs,
stars: black holes, stars: magnetars
\end{keywords} 
\section{Introduction}
\citet{A22,D22} discovered rapid fluctuations in the Rotation Measure (RM)
of FRB 20190520B and a reversal of its sign over several months.
\citet{D22} also reported rapid, apparently stochastic, variation of its
Dispersion Measure (DM).  In 113 bursts observed over 237 days DM ranged
between 1180 and 1230 pc-cm$^{-3}$.  67 bursts observed over two hours on
MJD 59373 showed DM ranging from 1190 to 1230 pc-cm$^{-3}$, suggesting a
rapidly varying and necessarily very compact near-source region contributing
0--40 pc-cm$^{-3}$.  The nearly constant remainder is plausibly contributed
by the intergalactic medium, Galactic plasma, and more distant (from the
source) portions of the host galaxy.

Typical observed bursts have peak flux densities $\sim 1\,$Jy and durations
1--2 ms, corresponding to a duty factor $\sim 10^{-5}$.  This could be the
result of isotropic emission $\sim 10^{-5}$ of the time, continual emission
into a wandering $\sim 10^{-4}\,$ sterad beam (readily interpreted as the
result of radiation by relativistic charge bunches with Lorentz factors
$\sim 100$), or some intermediate combination.  At the redshift $z = 0.241$
and observed bandwidth $\Delta \nu \sim 300\,$MHz the mean power is $\sim 3
\times 10^{36}\,$ergs/s.
\section{Bounds on Characteristic Time Scales}
\begin{figure}
	\centering
	\includegraphics[width=\columnwidth]{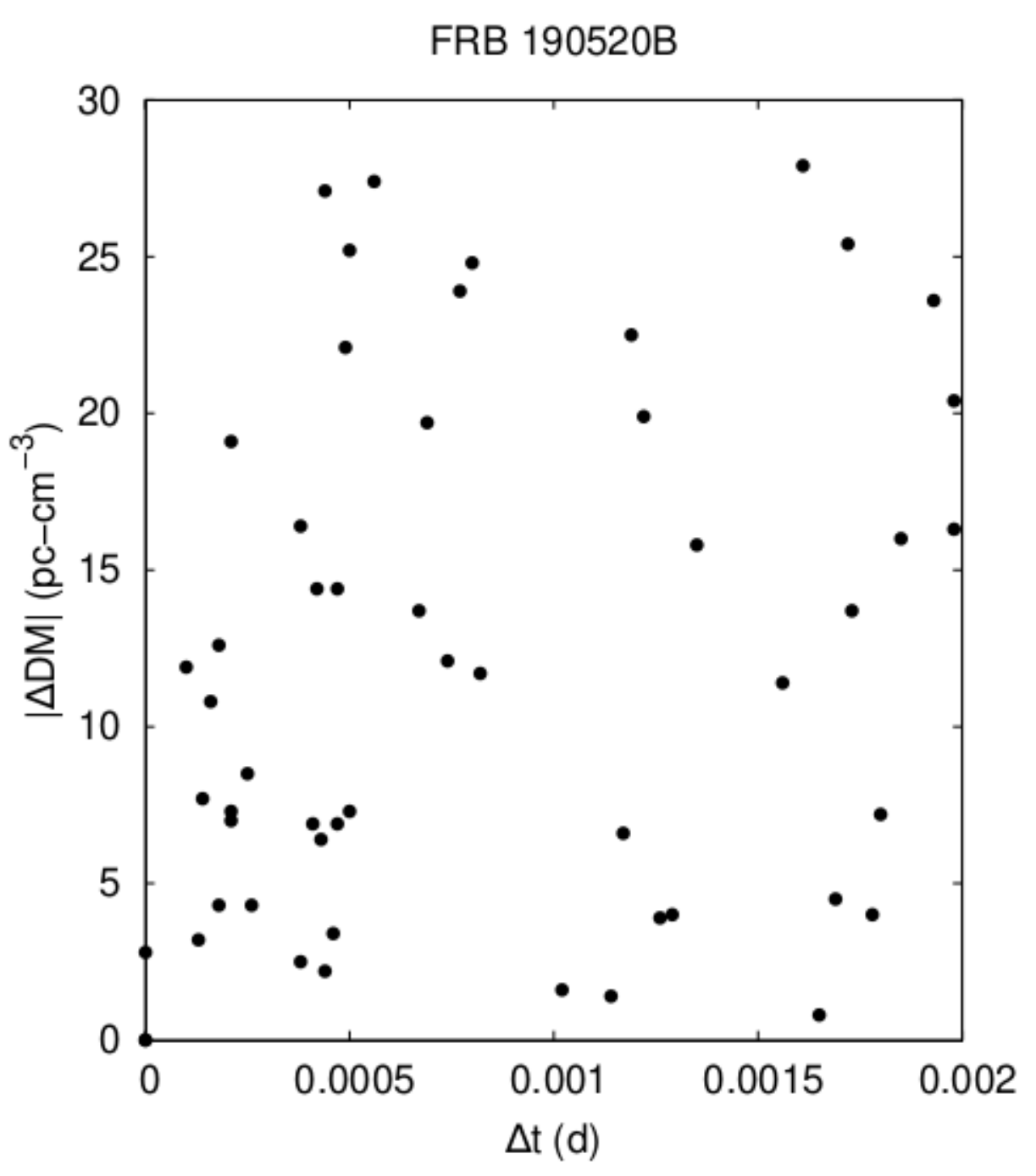}
	\caption{\label{DMvt} $\Delta|\text{DM}|$ {\it vs.\/} $\Delta t$ for
	intervals between successive bursts of FRB 20190520B on MJD 59373
	observed by \citet{D22}.  Uncertainties of $\Delta$DM are about $\pm
	0.6\,\text{pc-cm}^{-3}$, insignificant compared to the scatter of 
	$\Delta$DM.  Intervals with $\Delta t > 0.002\,$d are not shown.
	Two overlapping points at the origin (intervals of 5 ms and 23 ms
	between pairs of bursts with $\Delta \text{DM} = 0.0 \pm 0.6\,
	\text{pc-cm}^{-3}$) likely represent substructure of the same burst.
	The point on the ordinate with nominally significant $|\Delta
	\text{DM}| = 2.8 \pm 0.6\,$pc-cm$^{-3}$ and $\Delta t = 4 \times
	10^{-7}\,$d (35 ms) may also represent substructure.}
\end{figure}

Fig.~\ref{DMvt} shows the differences in |DM| between successive bursts
observed on MJD 59373 {\it vs.\/} their separation in time.

It is evident that there is no correlation of $|\Delta\text{DM}|$ with
$\Delta t$ for intervals as short as $1.0 \times 10^{-4}\,$d (8.64 s), the
shortest interval in the data of \citet{D22}, aside from three intervals
$\le 35\,$ms that may be attributable to burst substructure.  On the line of
sight to the FRB, likely surrounding it and causally associated with it, is
a turbulent region whose contribution to the DM varies from zero to 40
pc-cm$^{-3}$.

The data of Fig.~\ref{DMvt} set an upper bound $t \lesssim 10\,$s on the
characteristic time scale of this compact, near-source, turbulent region.  A
lower bound on the correlation time $\sim 35\,$ms may be estimated on the
basis of the small or zero $|\Delta \text{DM}|$ within burst substructure,
or between bursts separated by such small $\Delta t$.  The remaining 1190
pc-cm$^{-3}$ is attributed to the intergalactic medium, our Galaxy, and an
essentially unvarying (during the two hours of observation on MJD 59373, and
only slightly varying over the entire 237 day campaign), likely extended,
host galaxy region estimated to contribute about 900 pc-cm$^{-3}$
\citep{N21}.
\section{The Turbulent Region}
The turbulent region must be able to contribute $n_e R = \text{DM} \approx
40\,\text{pc-cm}^{-3} \approx 1.2 \times 10^{20}\,\text{cm}^{-2}$, where
$n_e$ is its characteristic electron density and $R$ its size (radius if
the FRB is at its center).

An upper bound on $n_e$ and lower bound on $R$ can be set from the fact that
\citet{D22} observed at frequencies 3000--3500\,GHz; propagation requires
that the the plasma frequency be lower than the frequency of observation.
Then
\begin{equation}
	\label{ne}
	n_e \le {\pi \nu^2 m_e \over e^2} \approx 1.1 \times 10^{11}\,
	\text{cm}^{-3},
\end{equation}
where $\nu \approx 3 \times 10^9\,$s$^{-1}$ is the frequency of the observed
radiation that has passed through the dispersing cloud.  There appears to be
a low-frequency cutoff around 3 GHz in the bursts observed by \citet{D22},
suggesting that Eq.~\ref{ne} may be an actual estimate of $n_e$ rather than
only a bound.

If the frequency of observation $\nu$ is close to the plasma frequency of
the dispersing cloud, then the dispersion delay is no longer quantitatively
proportional to $\nu^{-2}$, but has a more complex dependence on frequency:
\begin{equation}
	\label{DR}
	\Delta t = \int\!{d\ell \over c}\,{1 \over 2}{\omega_p^2 \over
	\omega^2} \left(1 + {3 \over 4}{\omega_p^2 \over \omega^2} +
	\cdots\right),
\end{equation}
where $\omega_p = \sqrt{4 \pi n_e e^2/m_e}$.  If the higher order term(s)
could be fit to the data, $n_e$ would be measured directly and $R$ inferred
unambiguously.

From the definition of DM
\begin{equation}
	\label{R1}
	R \gtrapprox {|\Delta\text{DM}| \over n_e}.
\end{equation}
The shortest time scale variations involve only $|\Delta \text{DM}| \approx
12\,\text{pc-cm}^{-3} \approx 3.5 \times 10^{19} \text{cm}^{-2}$; the
turbulent region is likely heterogeneous, with a greatest DM of 40
pc-cm$^{-3}$ but the most rapidly varying part contributing less.  Then,
using Eq.~\ref{ne},
\begin{equation}
	\label{R2}
	R \gtrapprox 3 \times 10^8\,\text{cm}.
\end{equation}

If the turbulent region is gravitationally bound to, or infalling into, a
mass $M$, then it is possible to bound this mass
\begin{equation}
	\label{GM}
	GM \sim {R^3 \over t^2} \gtrsim 4 \times 10^{23} {\text{cm}^3 \over
	\text{s}^2},
\end{equation}
or 
\begin{equation}
	\label{M}
	M \gtrsim 5 \times 10^{30}\,\text{g}.
\end{equation}
If $t$ is taken as the shortest burst interval $\sim 35\,$ms with nominally
significant non-zero $\Delta \text{DM}$ then the lower bound on $M$ would be
about five orders of magnitude greater, or $\sim 200 M_\odot$.  However,
this would be a slender reed on which to base such a remarkable inference.

Some of the earlier observations of \citet{N21} were made at about 1500 MHz,
but may have been taken at an epoch when the rapidly varying near-source
plasma was less dense.  \citet{N21} did not report a rapidly varying DM, so
the preceding analysis may not be applicable.  However, if it is applicable
and the $|\Delta \text{DM}|$ are comparable, the implied $n_e \lessapprox 3
\times 10^{10}\,$cm$^{-3}$, $R \gtrapprox 1.3 \times 10^9\,$cm (taking the
$|\Delta \text{DM}| = 12\,$pc-cm$^{-3}$ corresponding to the shortest
$\Delta t$, rather than the full range $|\Delta \text{DM}| =
40\,$pc-cm$^{-3}$), and the implied $M \gtrsim 4 \times 10^{32}\,$g. 

An independent bound can be derived from the requirement that the free-free
optical depth \citep{S62} be $\lesssim 1$.  Using Eq.~\ref{R1} to eliminate
$n_e$ in favor of DM,
\begin{equation}
	\label{RR}
	R > 5 \times 10^{10} \left({\text{DM} \over 12\,\text{pc-cm}^{-3}}
	\right)^2 T_6^{-3/2}\,\text{cm},
\end{equation}
where $T_6 \equiv T/(10^6\,\text{K})$.  Combining this with the causality
constraint $R < ct \approx 3 \times 10^{11}\,$cm yields
\begin{equation}
	\label{T6}
	T_6 > 0.3 \left({\text{DM} \over
	12\,\text{pc-cm}^{-3}}\right)^{4/3}.
\end{equation}

This is a physically possible condition, but indicates a hotter plasma
than the filaments of known supernova remnants.  However, the estimated
mean power of $3 \times 10^{36}\,$ergs/s, subject only to the distance
constraints of Eqs.~\ref{R2}, \ref{RR} and the causality limit $R \le ct$,
permits much higher energy density and temperature.  The alternative of
dispersion by the relativistic particles of a pulsar wind nebula fails
because they are insufficiently numerous to give significant dispersion,
and are also affected by the relativistic increase of effective mass.
\section{Magnetic Field}
The magnetic field in the dispersing region may be estimated:
\begin{equation}
	\label{Bp}
	|B_\parallel| \sim 1.23 {|\Delta \text{RM}| \over
	|\Delta \text{DM}|}\, \mu\text{G} \sim 300\,\mu\text{G},
\end{equation}
where RM is in the usual units of radians/m$^2$, DM is in pc-cm$^{-3}$,
and $B_\parallel$ is the electron density-weighted mean component along the
line of sight.

An alternative, but numerically similar, estimate may be obtained by
comparing the DM and RM of bursts 2 and 3 in Table 1 of \citet{D22}.  These
bursts occurred on MJD 59373 and the two methods of estimating RM agree
(burst 1 occurred on the same day, but the two estimates of RM disagree by
four times their formal error, an issue also for burst 5 on MJD 59400 and
burst 7 on MJD 59588).  Between bursts 2 and 3 $|\Delta \text{RM}| = 800 \pm
43$/m$^2$ and $|\Delta \text{DM}| = 1.8 \pm 0.5\,$pc-cm$^{-3}$, leading to
\begin{equation}
	|B_\parallel| = 546 \pm 164\,\mu\text{G}.
\end{equation}
The uncertainty chiefly results from the large fractional uncertainty in
$|\Delta \text{DM}|$.

The fact that DM varies coherently shows that there are not a large number
of independent regions contributing to the DM.  If this is also true for the
RM then Faraday rotation in regions with opposite signs of $B_\parallel$
does not efficiently cancel and Eq.~\ref{Bp} estimates the actual field
magnitude.  This is not proven because a region homogeneous in $n_e$ may have
subregions with cancelling $B_\parallel$; Eq.~\ref{Bp} is properly only a
lower bound on the field magnitude.

The field estimated in Eq.~\ref{Bp} is much smaller than the fields of 3--17
mG similarly estimated for the environment of FRB 121102 \citep{K21}.  This
results from the much larger $|\Delta \text{DM}|$ measured for FRB 190520B,
but why this should be so is unclear.
\section{Discussion}
The mass bound of Eqs.~\ref{GM}, \ref{M} isn't yet interesting, because it
is hard to imagine a FRB source with a mass less than the minimum mass of a
neutron star, about $1.4 M_\odot$.  However, future observations might well
produce lower estimates of the characteristic time $t$ or (if a rapidly
varying DM were observed at lower frequencies) larger estimates of the 
characteristic size $R$ from the requirement the plasma be dilute enough
to transmit radio waves of the observed frequency and yet provide the
measured $|\Delta \text{DM}|$.  These might provide useful constraints on
masses and models.

One conclusion can be drawn from the rapid variations of DM shown in
Fig.~\ref{DMvt} without modeling the dispersing region: It cannot be larger
than $ct \sim 3 \times 10^{11}\,$cm.  This excludes models in which the
dispersion is produced in a supernova remnant or other extended cloud.  It
also argues against models in which the FRB is produced far ($\gtrsim ct$)
from a central compact source unless the dispersing medium is heterogeneous
on scales $\ll ct$ and most of the dispersion is within $\ll ct$ of the
emission sites of the individual bursts (the medium must be extremely clumpy
with bursts and their dispersion produced in separate and uncorrelated
clumps).

The bound $R \lesssim ct$ sets a lower bound on $n_e$, independent of any
assumptions about the dynamics of the dispersing medium:
\begin{equation}
	\label{nect}
	n_e \gtrsim {|\Delta \text{DM}| \over ct} \sim 3 \times 10^8\,
	\text{cm}^{-3},
\end{equation}
consistent with Eq.~\ref{ne}.  This excludes the environment of a pulsar or
magnetar, swept clean of thermal plasma by its wind (relativistic plasma is
ineffective at dispersing radio emissions).  Supernova remnants have dense
filaments, but they would be expected to obscure the central object with a
low duty factor, while FRB 190520B appears to be behind dispersing matter
much of the time; its excess source region DM is broadly distributed 
(Fig.~\ref{DMvt}) from 0 to 40 pc-cm$^{-3}$ rather than having rare
excursions from low values.  Eq.~\ref{nect} may point to an accretion flow
\citep{K22}. 
\section*{Data Availability}
This theoretical study did not generate any new data.

\label{lastpage} 
\end{document}